%% file: main.tex
\documentclass[conference]{IEEEtran}
\IEEEoverridecommandlockouts
\usepackage[bottom=1.02in,top=0.71in,left=0.7in,right=0.7in]{geometry}
\usepackage{amsmath}
\usepackage{amsfonts}
\usepackage{amssymb}
\usepackage[ruled,vlined]{algorithm2e}
\usepackage{caption}
\usepackage{mathtools}  
\usepackage{amsthm}
\usepackage[]{algorithm2e}
\usepackage{booktabs}
\usepackage{subfigure}
\usepackage{tabulary}
\usepackage{multirow}
\usepackage{footnote}
\usepackage[export]{adjustbox}
\usepackage{booktabs}
\usepackage[justification=centering]{caption}
\usepackage{comment}

\usepackage{cases}
\usepackage{graphicx}
\usepackage{cite}
\usepackage{multicol}
\usepackage{xcolor}
\usepackage{graphicx}
\graphicspath{{./}{figures/}}
\newcommand\numeq[1]%
  {\stackrel{\scriptscriptstyle(\mkern-1.5mu#1\mkern-1.5mu)}{=}}
\usepackage{stfloats}
\usepackage{cuted}
\usepackage{lipsum}  

\usepackage{dsfont}
\usepackage{amsmath,amssymb}

\newtheorem{Prob}{\textbf{Problem}}

\usepackage{optidef}
\usepackage{amsmath}
\title{LLP-V2X: Low Latency-Power Vehicular Networking Towards 6G V2X}

\author{
    \IEEEauthorblockN{
        Zhaoyu Liu\IEEEauthorrefmark{1}, 
        Liu Cao\IEEEauthorrefmark{1},
        Lyutianyang Zhang\IEEEauthorrefmark{2},
        Dongyu Wei\IEEEauthorrefmark{3},
        Ye Hu\IEEEauthorrefmark{4},
        Weizheng Wang \IEEEauthorrefmark{5}\\
    }
    \IEEEauthorblockA{
        \IEEEauthorrefmark{1}City University of Hong Kong (Dongguan), Dongguan, China\\
         \IEEEauthorrefmark{2} The School of Microelectronics and Communication Engineering, Chongqing University, Chongqing, China\\
        \IEEEauthorrefmark{3} Department of Electrical and Computer Engineering, University of Miami, FL, USA\\
         \IEEEauthorrefmark{4} Department of Industrial and Systems Engineering, University of Miami, FL, USA\\
        \IEEEauthorrefmark{5} 
        Department of Computer Science, City University of Hong Kong, Hong Kong SAR\\
        Emails: \{72515198, liu.cao\}@cityu-dg.edu.cn, zhanglyutianyang@cqu.edu.cn, \{dongyu.wei, yehu\}@miami.edu, \\weizheng.wang@ieee.org
    }
    \vspace{-0.5cm}
\thanks{This work was supported by the Youth Innovation Talent Project of Guangdong Provincial Universities (Grant No. 2025KQNCX17).}
}

\begin{document}

\maketitle
\thispagestyle{empty}
\begin{abstract}

The trade-off between energy and latency budgets is becoming significant due to the more stringent QoS requirements in 6G vehicular networks. However, comprehensively studying the trade-off between energy and latency budgets for 6G vehicular network with new Vehicle-to-Everything (V2X) features is still under-explored. This paper proposes a novel multi-hop, multi-path vehicular networking that jointly optimizes vehicular traffic splitting across candidate routes and per-link transmit power to achieve low-latency and low-power communications. Afterwards, we formalize two complementary problem formulations (minimum latency and minimum power) based on the proposed 6G V2X architecture and provide sufficient conditions. The performance of the proposed scheme is evaluated via well-designed simulations. Based on these theories, we design algorithm (LLP MHMP Scheduler) that switches on demand between a fixed-power minimum-latency mode and a fixed-latency minimum-power mode.

\end{abstract}


\input{introduction}
\input{sys_model}

\input{simulation}

\input{conclusion}

\bibliographystyle{IEEEtran}
\bibliography{ref}
\end{document}

%% file: introduction.tex
\section{Introduction}
\label{sec:intro}

The rapid commercialization of 5G-enabled Internet of Vehicles (IoV)---with high throughput, low latency, and massive connectivity---has moved services from infotainment to safety-adjacent intelligent driving (e.g., in-cabin cloud gaming, multimodal AI assistants, low-latency VR/AR, remote functions). Yet 5G is insufficient for data- and coordination-intensive tasks (higher-level autonomy, group collaboration): sustained high-rate perception streams (e.g., LiDAR/HD video) are limited by effective payload and congestion under dense access; uneven coverage, fading, and multi-user contention depress practical rates; and high mobility inflates end-to-end delay, hindering millisecond-scale cooperative control (platooning, coordinated obstacle avoidance). Architecturally, emphasis on point-to-point V2I/V2C/V2V links with little native multi-hop self-organization constrains wide-area sharing and distributed decision-making across vehicle cohorts and road segments, leaving system-level gains from collective optimization largely unrealized.

By contrast, 6G IoV will fuse ultra-high-resolution LiDAR, multispectral/thermal imaging, and environmental sensing into massive, high-dimensional multimodal streams\cite{wang2021green}. Realizing \emph{collective intelligence}---collaborative perception, predictive model exchange, coordinated control---demands real-time V2V/V2I dissemination beyond simple status beacons and stresses on-board ECUs. Accordingly, 6G shifts toward decentralized, self-organizing, multi-hop meshes where vehicles act as intelligent relays. While this architecture enables large-scale cooperation, it amplifies communication and computation loads, making energy consumption a primary feasibility bottleneck in dense, highly dynamic vehicular meshes.

Recent advances in multi-path and multi-hop transmission have been widely leveraged to enhance the reliability and efficiency of vehicular networks\cite{diniesh2024addressing,wang2024amtos,10578028,11096953,cao2022optimize,cao2022resource,hu2024multi,LC2025IoTJ}. However, current 5G systems for collaborative driving remain constrained by small-packet capacity limits (e.g., transport block (TB) / MAC PDU sizes) and a predominantly point-to-point link abstraction, which together make it difficult to support fleet-oriented massive sensor-data sharing and multi-hop cooperation. Existing enhancements (e.g., packet-size-aware scheduling, asymmetric dual-mode frameworks, and reinforcement-learning-based resource allocation) largely stay at the single-hop or physical/link-layer level and lack joint guarantees for \emph{end-to-end} latency, jitter, and ultra-high reliability for ``large packets'' under dynamic topologies, as well as a global multi-path scheduling perspective. Although NOMA/OFDMA and AoI-driven spectrum allocation can reduce retransmissions and control overhead—thus lowering per-node power under the same traffic load—these methods are still \emph{local optimizations} from a resource-allocation viewpoint\cite{newaz2021optimizing}. For vehicle-dense 6G V2X scenarios, there is a pressing need for \emph{system-level}, \emph{cross-layer}, and \emph{multi-node concurrent} low-power designs to achieve wide-area collaboration across vehicle groups and road segments while optimizing energy efficiency.In terms of reducing latency, the literature\cite{zhou2022delay} proposes an evolutionary Dijkstra algorithm that can find routing solutions with low complexity to minimize latency, and a larger number of FD nodes is more conducive to reducing latency, ultimately achieving the minimum end-to-end latency. But it did not take into account multi-path transmission and power optimization.


Motivated by the above limitations, we propose a novel
6G V2X architecture with optimization methods to achieve the low power and low latency V2X communications. The main contributions are summarized as follows:
\begin{itemize}
\item We extend the V2X communication architecture from the originally standardized single-hop \footnote{At most two-hop transmission with one-relay vehicle is supported for 5G V2X per the 3GPP V2X standard \cite{3gpp.37.985}.} single-path for 5G V2X \cite{3gpp.38.885,Garcia2021NRV2X} to the multi-hop multi-path for 6G V2X.
\item We proposed a dynamic traffic allocation policy that steers and splits multiple service types of traffic simultaneously to achieve the minimum latency or power performance driven by their QoS requirements, i.e., the energy or latency budget. We further propose an adaptive low-Latency and low-power scheduler based on the trade-off between the latency and energy budget.
\end{itemize}


The rest of this paper is organized as follows: Sec \ref{sec:sys_arc}  introduces the system architecture. The system model is illustrated in Sec \ref{sec:sys_model}. In Sec \ref{sec:sim}, we compared the proposed multi-path multi-hop solution with the baseline solution in terms of average delay and power through simulation. \ref{sec:con} draws the conclusions for this paper. 




%% file: sys_model.tex
\section{Proposed Multi-Hop Multi-Path V2X Communication Architecture}
\label{sec:sys_arc}

\begin{figure}[t]
    \centering
\includegraphics[width=.48\textwidth]{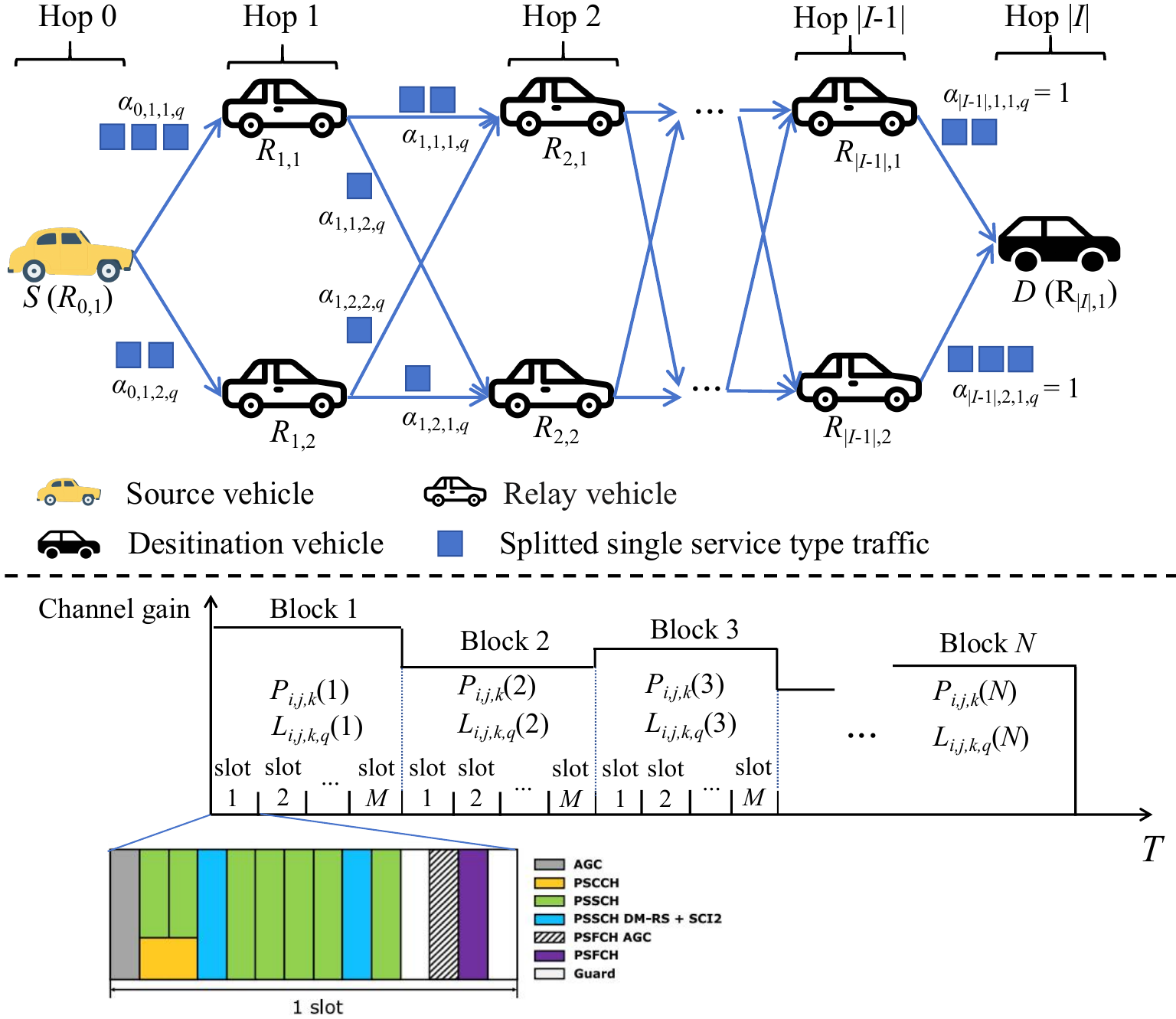}
    \caption{Proposed Multi-hop Multi-path Architecture for Multi-type V2X Services of Traffic.}
    \label{fig:E2Esysarch}
    \vspace{-0.65cm}
\end{figure}

Let us consider an End-to-end (E2E) multi-hop multi-path V2X sidelink communication architecture, as is shown in Fig. \ref{fig:E2Esysarch}. A single source vehicle \(S\) transmits $|Q|$ service types of V2X traffic to a destination vehicle \(D\) through an $|I|$-hop multi-path vehicular network where the hop $1 \leq i\in I \leq |I|-1$ is pre-configured to contain a fixed $|J| = 2$ relay vehicles while the hop $0$ and hop $|I|-1$ only contains \(S\) and \(D\), respectively. We denote the $j$th relay vehicle ($j\in J = \{1,2\} $) in the $i$th hop as \(R_{i,j}\). As a result, each relay vehicle in the $i$th ($i \in \{0,1,2,...,|I|-2\}$) hop transmits traffic over the $|K| = |J|= 2$ paths to the pre-configured dual relay vehicles in the $i+1$th hop while each relay vehicle in the $|I|-1$th hop transmits traffic over the single path to the last hop $|I|$ containing the single destination vehicle. As each hop $ 1\leq i\in I \leq |I|-1$ contains a fixed $|J| = 2$ relay vehicles, the number of \textbf{E2E} (i.e., from \(S\) to \(D\)) paths should be $2^{|I|-1}$.

In this work, we assume a block fading channel instance with a fixed time interval $T_c$ (i.e., coherence time) consisting of $M$ consecutive 5G New Radio slots \footnote{The slot duration is determined by 5G numerology $\mu = \{0,1,2,3,4,5\}$ that determines the subcarrier spacing (SCS) of $15 \times 2^{u} \text{kHz}$ in the 3GPP standard \cite{3gpp.36.321}.} and different channel instances are assumed to be independent and identically distributed (i.i.d.). Thus, we consider a time-block system with a fixed block duration $T_c$, and each relay vehicle makes joint optimization decisions in terms of the split of traffic for different traffic services and the allocation of transmit power to dual paths at the beginning of each block $n \in N = \{1,2,...,|N|\}$. Meanwhile, we assume the optimization decisions determined at the beginning of each block will last for a whole block duration. Note that our proposed multi-hop multi-path V2X communication procedures can be also well compatible with the existing 3GPP V2X standard: As Fig.1 shows, each of the $i$th hop's relay vehicles is able to estimate the per-path link quality based on the feedback from the $i+1$th hop's vehicles via the sidelink feedback channel PSFCH, and then transmits the information of scheduling decisions on the sidelink control channel PSCCH and sidelink shared channel PSCCH with DMRS and SCI format 2 while the traffic payload is transmitted on the sidelink shared channel PSSCH \cite{Garcia2021NRV2X}. 

Under the above system architecture setup, let $0 \le \alpha_{i,j,k,q}(n) \le 1$ be the splitted traffic ratio of service type $q\in Q$ from the relay vehicle \(R_{i,j}\) routed to the next hop relay vehicle \(R_{i+1,j}\) over the $k$th path ($k\in K$). Meanwhile, let $P_{i,j,k}(n)$ be the allocated power from the relay vehicle \(R_{i,j}\) over the $k$th path. Then each relay vehicle makes joint scheduling decisions in terms of $\alpha_{i,j,k,q}(n)$ and $P_{i,j,k}(n)$ at the beginning of block $n$. Then the resulting estimated instant latency for the service type $q\in Q$ of traffic during the block $n$ is denoted as $L_{i,j,k,q}(n)$. Note that different service types of traffic have different optimization decisions made by each relay vehicle within a block due to different traffic characteristics and wireless environments.

\section{Problem Formulation}
\label{sec:sys_model}
In this section, we provide problem formulations based on the E2E multi-hop multi-path V2X architecture. Assume the number of arrival packets for service type $q$ is a random variable $\lambda_{i,k,q}$ following a Poisson distribution with the given mean $E[\lambda_{i,k,q}] = \overline{\lambda_{i,k,q}}$. The packet size for the service type $q$ is $M_{i,k,q}$ (number of bits), while it is assumed that the packet size of all packets belonging to one service type is identical. Meanwhile, we denote the number of packets for service type $q$ at UE queue at the beginning of each time slot as $n_{i,j,q}$, which also follows a Poisson distribution with the given mean $E[n_{i,j,q}] = \overline{n_{i,j,q}}$. Therefore, the estimated instant latency during block $n$ is given by
\begin{align}\label{eqn: UulatencyPath1}
L_{i,j,k,q}(n)= \frac{\alpha_{i,j,k,q}(n)(\lambda_{i,k,q}(n)+n_{i,k,q})M_{i,k,q}(n)}{D_{i,j,k}(n)},  
\end{align}
where $D_{i,j,k}(n)$ is function of transmit power $P_{i,j,k}(n)$ allocated to the $k$th path:
\begin{align}\label{eqn: R1}
{D_{i,j,k}}(n) = B_{i,j}\mathrm{log_2}\left(1+\frac{P_{i,j,k}(n) - P_{i,j,k}^{Loss} }{B_{i,j} N_o}\right),
\end{align}
where $B_{i,j}$ is the bandwidth allocated to $R_{i,j}$. We assume all relay vehicles share the same equal bandwidth. The pathloss and shadowing effects on the $k$th path  follows the rule of large scale fading which is specified in the 3GPP standard \cite{3gpp.38.901}. $N_o$ is the noise power spectral density (PSD). As a result, the estimated instant latency of the $b$th \textbf{E2E} path from $S$ to $D$ for service type $q\in Q$ during block $n$ is given by 
\begin{equation}\label{eq:Ti}\begin{aligned}
u_{q,b}(n) &= \sum_{i\in I} L_{i,j,k,q}(n), 
\end{aligned}\end{equation}
where $b \in B =\{1,2,...,2^{|I|-1}\}$. 

Our objective is to achieve the E2E low latency or the low power within the proposed multi-hop multi-path V2X architecture for all service types of traffic. Thus the corresponding optimization problems are formulated as below:

\begin{Prob}[E2E Low-latency Multi-hop Multi-path Optimization]
\label{Minimum Latency}
\begin{equation}
\begin{aligned}
& \underset{\alpha_{i,j,k,q}(n),P_{i,j,k}(n)}{\mathrm{argmin}} 
 \sum_{q \in Q} \mathrm{max} \ u_{q,b}(n)  \\
\mathrm{s.t.} ~~~~&\mathrm{C1}: 0\leq u_{q,b}(n) \leq L_{q}^{max}, q\in Q, b \in B\\
&\mathrm{C2}: 0 \leq \alpha_{i,j,k,q}(n) \leq 1, i\in I,j\in J,\ k \in K,\ q\in Q,\\
&\mathrm{C3}:   \sum_{k \in K} \alpha_{i,j,k,q}(n) = 1, i\in I,j\in J, q\in Q,\\
& \mathrm{C4}: 0 \leq P_{i,j,k}(n) \leq P_{tot}, i\in I,j\in J,\ k\in K,\\
& \mathrm{C5}: \sum_{k \in K}P_{i,j,k}(n) =  P_{tot}, i\in I,j\in J,\\
\label{p:mian}
\end{aligned}
\end{equation}
\end{Prob}

As the instant latency of each service type of traffic during block $n$ is determined by the maximum instant latency among the $2^{|I|-1}$ paths, we thereby use $\mathrm{max} \ u_{q,b}(n)$ to represent the estimated \textbf{instant} latency for service type $q$ of traffic during block $n$. The constraint C1 indicates the estimated instant latency of service type $q$ on each path should not exceed its latency budget $L_q^{max}$. C2 defines the range of traffic ratio on each path while C3 defines the sum of splitted traffic ratio from a relay vehicle should be 1. C4 defines the range of the transmit power from a relay vehicle allocated to each path, and C5 indicates that the sum of allocated transmit power should be the total transmit power $P_{tot}$ of a vehicle. We assume all vehicles have the same transmit power $P_{tot}$. 

As a result, the estimated \textbf{average} latency for the service type $q$ of traffic over all $N$ blocks is given by
\begin{equation}
\bar L_q=\frac{1}{N}\sum_{n=1}^{N}u_{q,b}^*(n),
\end{equation}
where $u_{q,b}^*(n)$ is the instant latency for the service type $q$ with the optimized $\alpha_{i,j,k,q}^*(n)$ and $P_{i,j,k}^*(n)$ during the block $n$. 

\begin{Prob}[Low-power Multi-hop Multi-path Optimization]
\label{Minimum Power}
\begin{equation}
\begin{aligned}
& \underset{\alpha_{i,j,k,q}(n),P_{i,j,k}(n)}{\mathrm{argmin}}  P(n)= \sum_{i \in I, j\in J, k \in K,} P_{i,j,k}(n)  \\
\mathrm{s.t.} ~~~~&\mathrm{C1}:0 \leq P_{i,j,k}(n) \leq P_{tot}, i\in I,j\in J,\ k\in K,\\
&\mathrm{C2}: 0 \leq \alpha_{i,j,k,q}(n) \leq 1, i\in I,j\in J,\ k\in K,\ q\in Q,\\
&\mathrm{C3}:   \sum_{k \in K} \alpha_{i,j,k,q}(n) = 1, i\in I,j\in J, q\in Q,\\
& \mathrm{C4}: \mathrm{max}(u_{q,b}(n)) = L_{q}^{max}, q\in Q, b \in B,\\
\label{p:mian1}
\end{aligned}
\end{equation}
\end{Prob}


\noindent where C4 indicates the estimated instant latency for service type $q$ of traffic during block $n$ is set to the its latency budget $L_q^{max}$ in order to achieve the minimum transmit power of all transmitting vehicles. 

Therefore, the estimated \textbf{average} power of the whole vehicular network over all $N$ blocks is given by
\begin{equation}
\bar P=\frac{1}{N}\sum_{n=1}^{N}P^*(n),
\end{equation}
where $P^*(n)$ is the instant power with the optimized $\alpha_{i,j,k,q}^*(n)$ and $P_{i,j,k}^*(n)$ during the block $n$.

%% file: simulation.tex
\section{Simulation}
\label{sec:sim}

In this section, we first conducted the simulation for a \textbf{3-hop} scenario in Matlab to test our proposed optimization framework in terms of E2E latency as well as the power assumption. We then extend the simulation with more hops to show the impact of number of hops on the latency savings and power savings.  Meanwhile, we also propose three baseline solutions to justify the effectiveness of the proposed optimization framework, which are described as follows:
\begin{itemize}
\item \textbf{Single-path (SP):} The vehicle fixes one path for the entire horizon.In terms of achieving minimum latency in a single path,we pick the globally best fixed path among the candidates. While transmitting on a single fixed path with the minimum power that satisfies the maximum latency requirement.
\item \textbf{Path-selection (PS):} At each time interval, the vehicle dynamically selects from the candidate end-to-end paths.In terms of achieving the low latency(LL),the LL PS scheme 1 is to select the best path for each layer without performing general multi-path segmentation. The LL PS scheme 2 is to select two paths among the four paths for simultaneous transmission. In order to achieve the low power(LP), LP PS scheme 1 is selecting two optimal paths transmission under fixed delay requirements, and power is centrally allocated between these two paths to reduce total power. While LP PS scheme 2 is selecting the Non-cross topology.
\item \textbf{Multi-hop Multi-path (MHMP):} On multi-hop, multi-path topologies the vehicle is allowed, at each time interval, to split traffic across multiple end-to-end paths and to jointly allocate per-hop transmit power. Latency-oriented runs minimize end-to-end latency (LL MHMP); power-oriented runs minimize total transmit power subject to the fixed-latency requirement (LP MHMP).
\end{itemize}

In our simulation scenario, We consider vehicles with a communication coverage radius of $r=50$ m and assume that all nodes remain in the overlapping coverage area throughout the simulation, so connectivity is guaranteed. The main simulation setup is summarized in Table~\ref{tab:sim_para.} and Table~\ref{tab:sim_para.}.  

\begin{table}[t]
 \centering
 \caption{\small{Main simulation parameters.}}
 \label{tab:sim_para.}
 \resizebox{.50\textwidth}{!}{
 \begin{tabular}{ |c|c| }
 \hline
 \textbf{Parameter} & \textbf{Value} \\
 \hline
 Service types, $q$ & \{1, 2\} \\
 \hline
 Packet size, $M_q$ (bytes) & Traffic 1/2: 250/100 \\
 \hline
 Average packet arrival rate (pkts/s) & From traces $\{\lambda_{i,j,k}\}$ (traffic 1, traffic 2) \\
 \hline
 Average number of queuing packets, $n_i$ & Trace–driven (slot-wise measured) \\
 \hline
 Latency constraint, $L_q^{\max}$ (second) & 0.03 (30\,ms) for fixed-latency runs \\
 \hline
 Noise PSD, $N_0$ (dBm/Hz) & $-174$ \\
 \hline
 Center frequency band (GHz) & 5.9 \\
 \hline
 Total transmit power, $P_{\mathrm{tot}}$ (dBm) & 23 \\
 \hline
 Total bandwidth (MHz) & 100 \\
 \hline
 Pathloss and Shadowing & 3GPP TR 38.901 UMa NLoS \cite{3gpp.38.901} \\
 \hline
 Speed of relay vehicles (m/s) & 1 \\
 \hline
 Block duration, $T_c$ (s) & 0.5 \\
 \hline
 Numerology, $\mu$ (ms) & 0 \\
 \hline
 Simulation time (s) & 300 \\
 \hline
 CPU & Intel Core i7 @ 2.6\,GHz \\
 \hline
 \end{tabular}}
\end{table}


The proposed optimization framework is implemented in Matlab and solved with \textsc{CVX} on a per-block basis. We evaluate two objectives: (i) minimizing latency and (ii) minimizing power under three schemes. For a fair comparison, the following resource rules are enforced. Under SP and PS, the vehicle assigns the entire transmit power budget $P_{\mathrm{tot}}$ to the chosen path. In the MHMP, traffic is split across the available E2E paths and per-hop powers are jointly optimized subject to node budgets; in contrast, SP does not perform traffic splitting and PS only can select up to two paths to perform traffic diversion. For latency-oriented evaluation, we report per-block latency distributions as well as long-term average latency for each traffic type; for power-oriented evaluation, we measure the total transmit power under the same latency constraints. Unless otherwise stated, random seeds are fixed, and results are averaged over multiple Monte Carlo runs to mitigate stochastic variability.

\begin{figure*}[t]
\begin{minipage}[t]{0.24\linewidth}
\centering
 \subfigure[Instant latency CDF of traffic 1.]
{\includegraphics[width=1.7in]{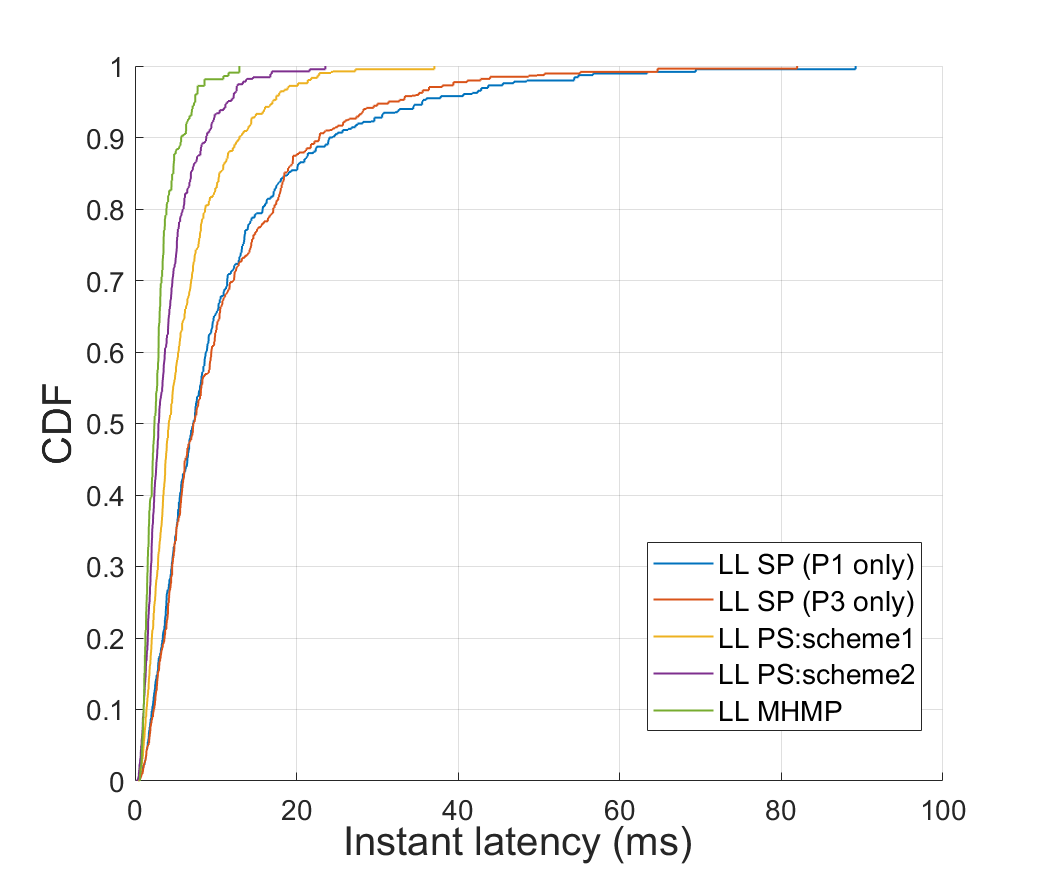}}
\label{fig:2a}
\end{minipage}
\begin{minipage}[t]{0.24\linewidth}
\centering
\subfigure[Instant latency CDF of traffic 2.]{
\includegraphics[width=1.6in]{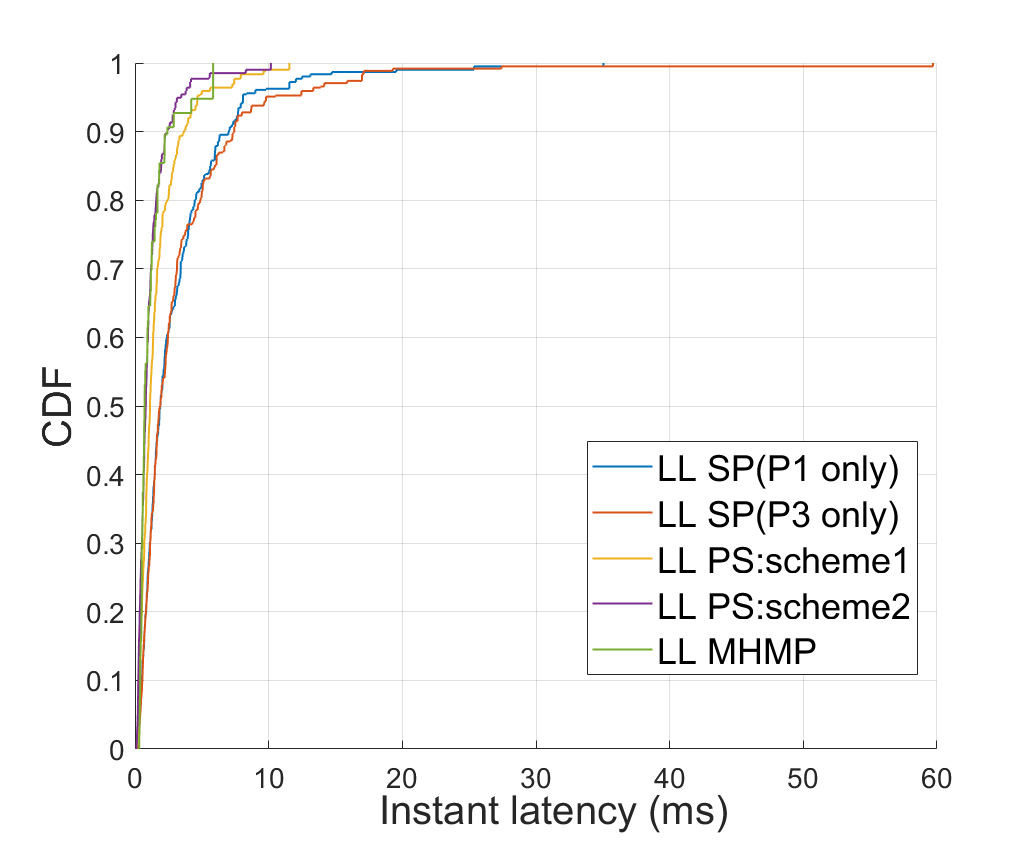}}
\label{fig:2b}
\end{minipage}%
\begin{minipage}[t]{0.24\linewidth}
\centering
\subfigure[Optimized traffic ratio.]{
\includegraphics[width=1.6in]{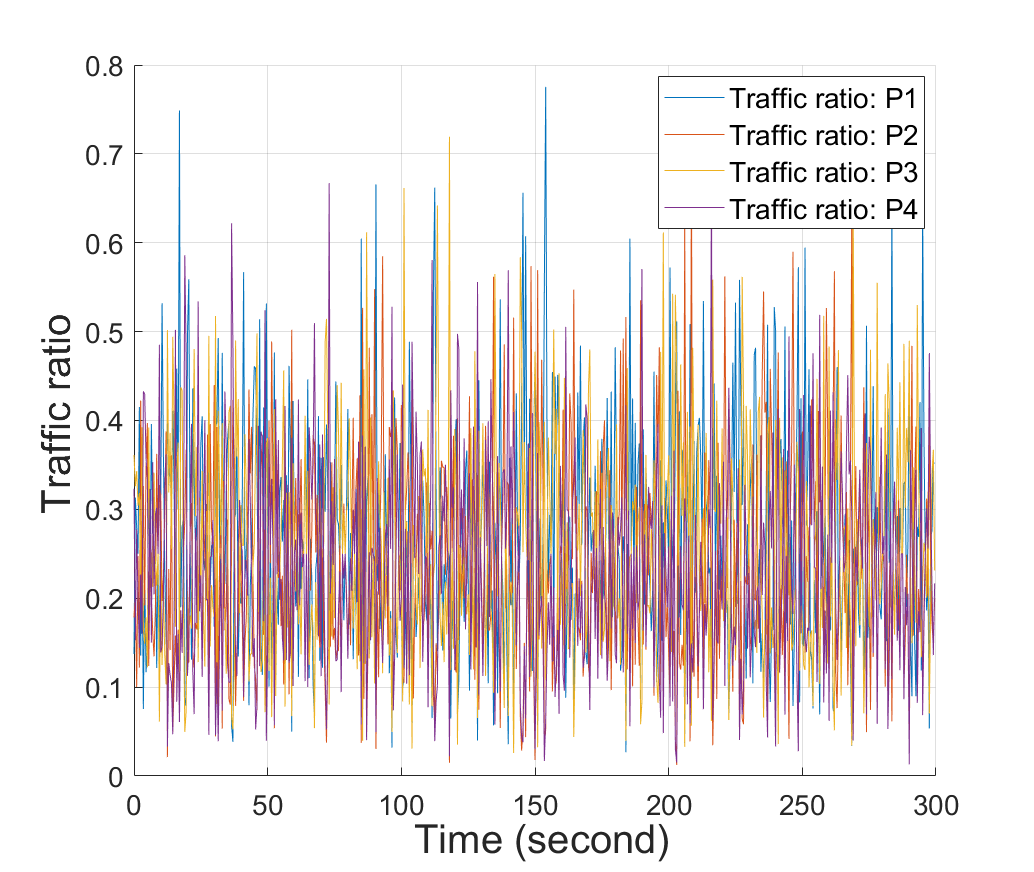}}
\label{fig:2c}
\end{minipage}
\begin{minipage}[t]{0.24\linewidth}
\centering
\subfigure[Optimized transmit power.]{
\includegraphics[width=1.6in]{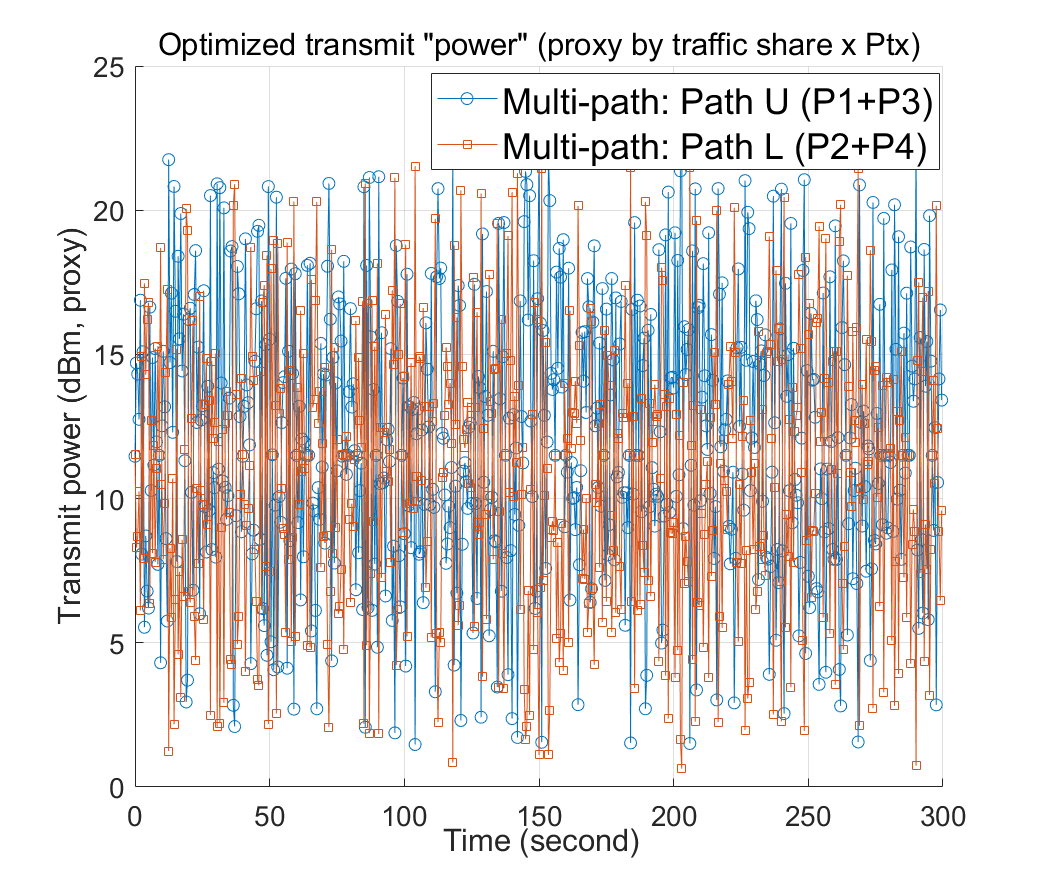}}
\label{fig:2d}
\end{minipage}
\caption{E2E Low Latency Aware Multi-Hop Multi-path Performance: A Three-hop Scenario.}
\label{fig:2}
\end{figure*}

\begin{figure*}[t]
\begin{minipage}[t]{0.24\linewidth}
\centering
 \subfigure[Instant latency CDF of traffic 1.]
{\includegraphics[width=1.6in]{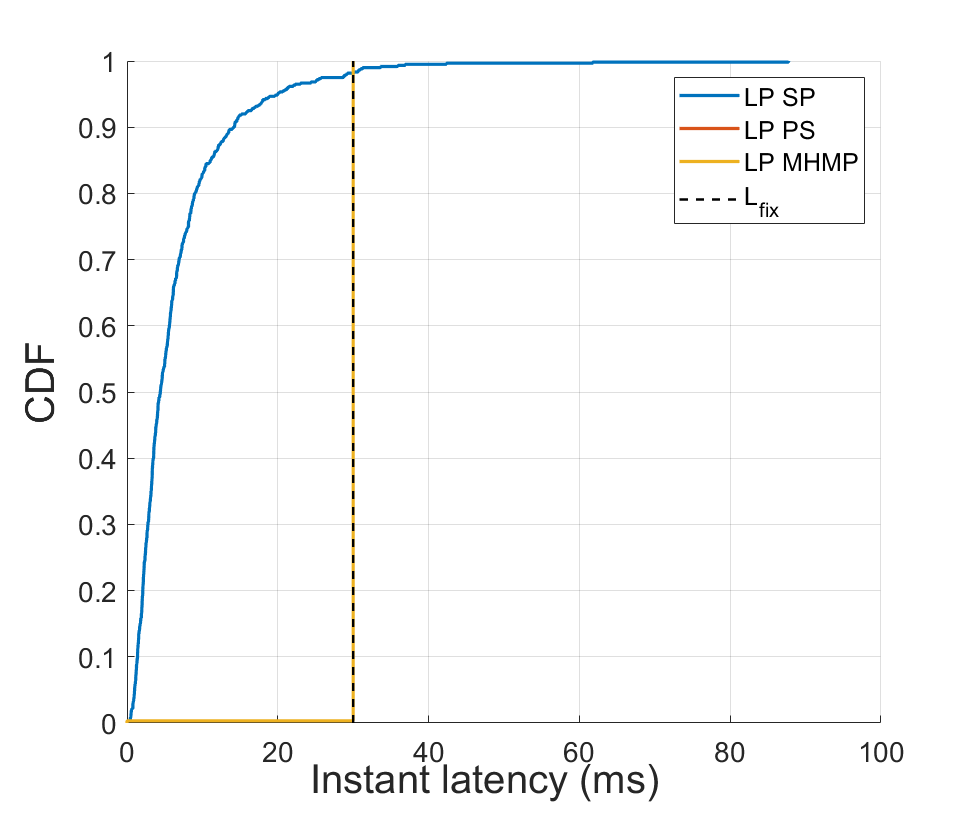}}
\label{fig:Latency CDF}
\end{minipage}
\begin{minipage}[t]{0.24\linewidth}
\centering
\subfigure[Optimized power comparison: MHMP vs. PS Scheme 1.]{
\includegraphics[width=1.55in]{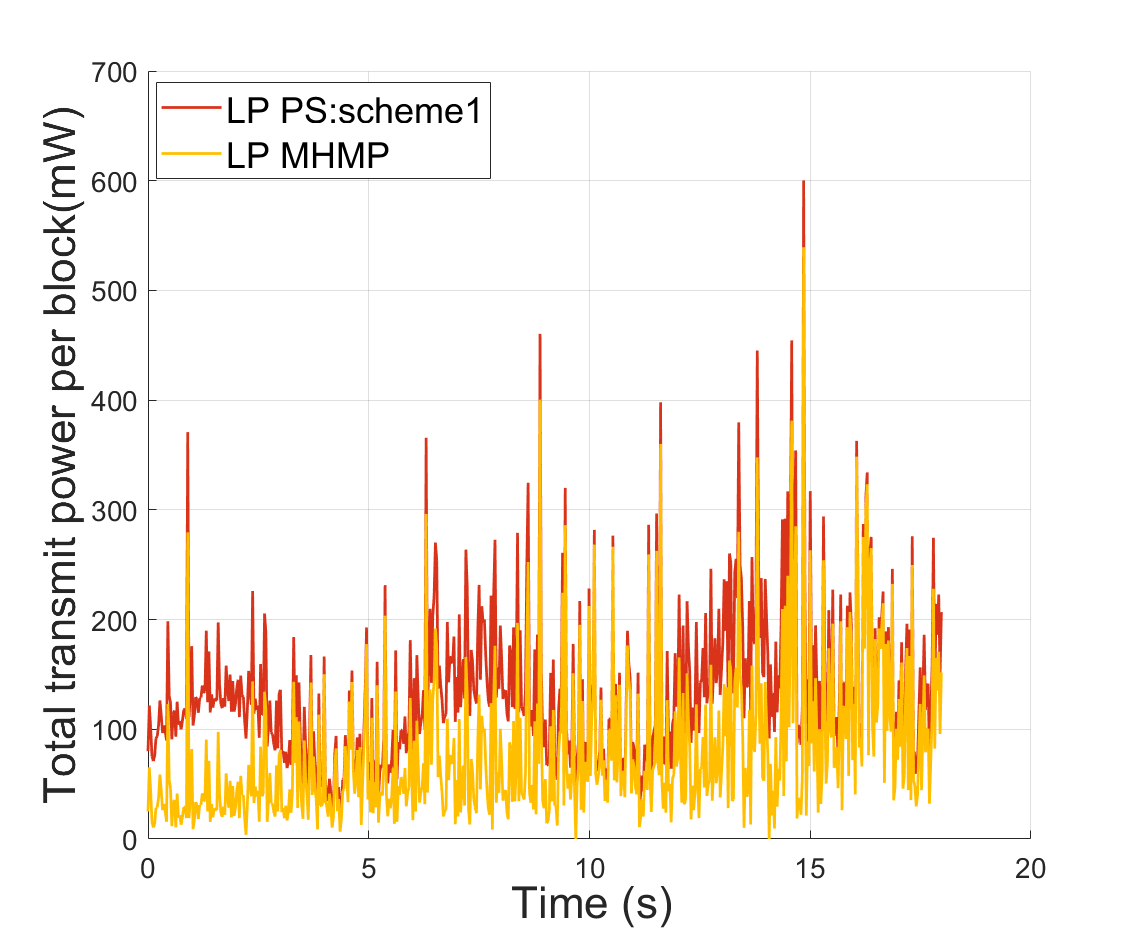}}
\label{fig:final}
\end{minipage}%
\begin{minipage}[t]{0.24\linewidth}
\centering
\subfigure[Optimized power comparison: MHMP vs. PS Scheme 2.]{
\includegraphics[width=1.55in]{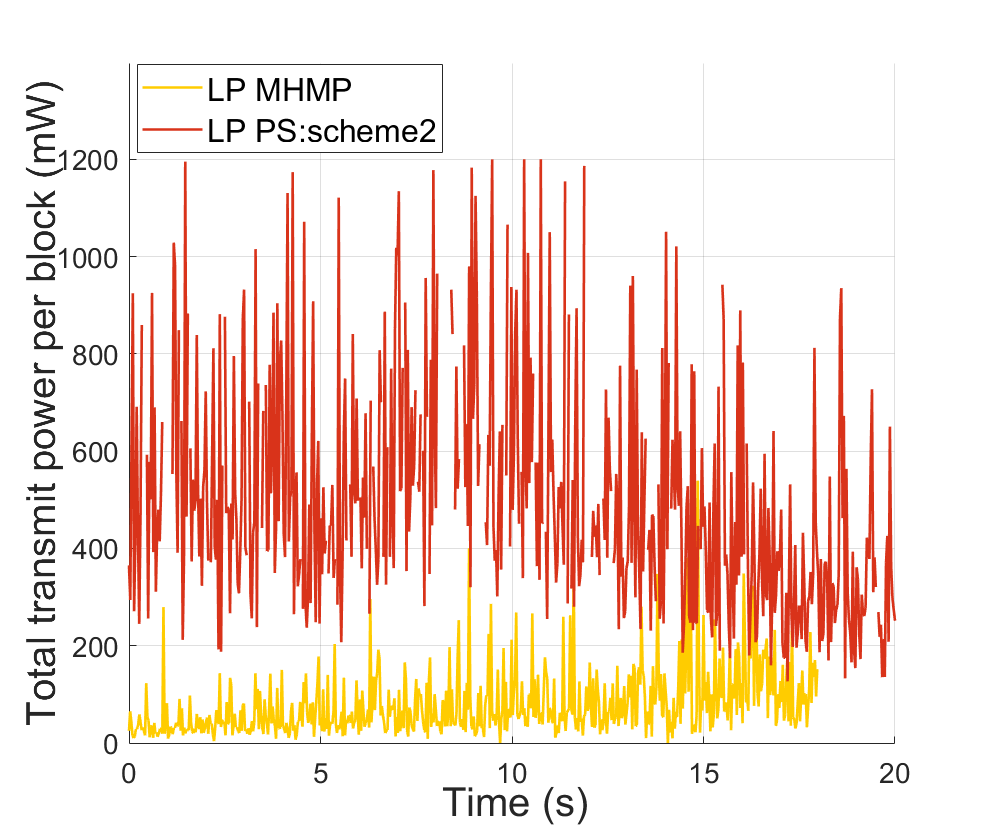}}
\label{fig:red}
\end{minipage}
\begin{minipage}[t]{0.24\linewidth}
\centering
\subfigure[Optimized power comparison: MHMP vs. SP.]{
\includegraphics[width=1.55in]{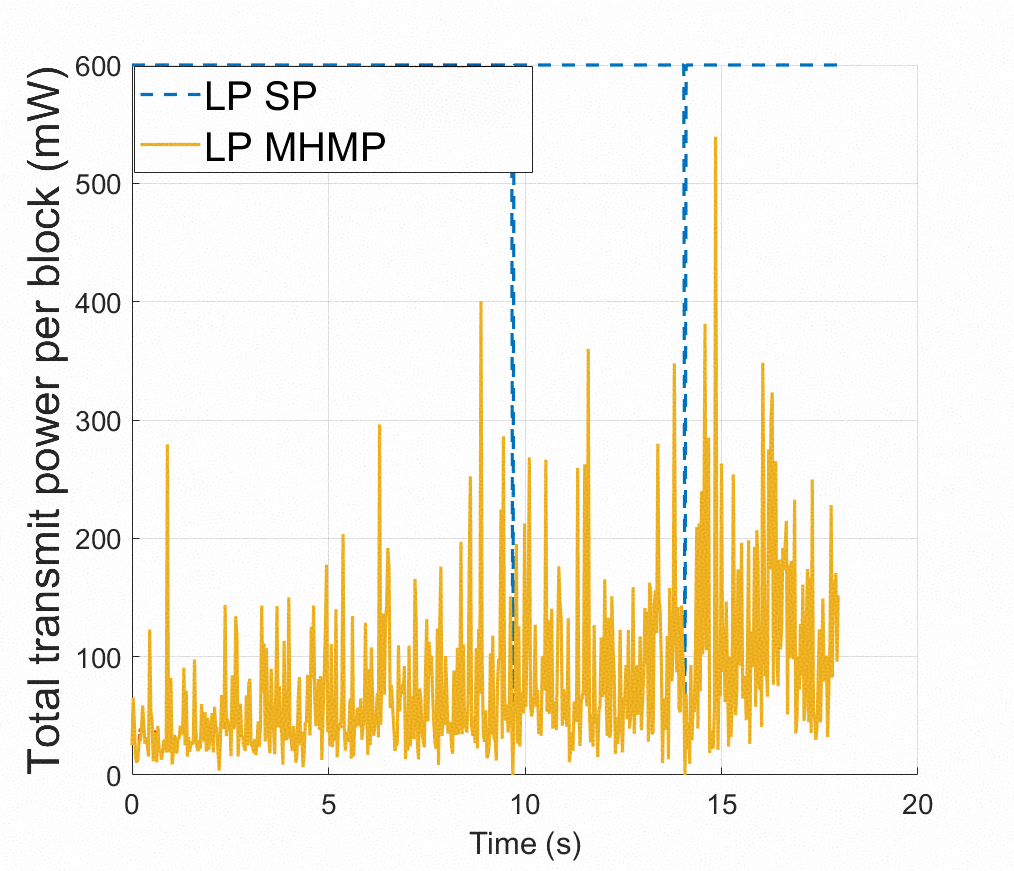}}
\label{fig:Total power_new}
\end{minipage}
\caption{Low Power Aware Multi-Hop Multi-path Performance: : A Three-hop Scenario.}
\label{fig:3}
\vspace{-0.5cm}
\end{figure*}

\subsection{Multi-Hop Multi-Path performance analysis}
We first discuss the optimization of latency and compare the performance of the proposed multi-path solution based on the baseline. In this case, the distance between each vehicle is set to 50 meters.

\begingroup
\setlength{\tabcolsep}{4pt}       
\begin{table}[ht]
  \centering
  \caption{\small{Average latency (ms) and average power (mW).}}
  \label{tab:avgLatPwrS1}
  \scriptsize
  \resizebox{.48\textwidth}{!}{%
  \begin{tabular}{|l|l|c|c|c|c|c|}
    \hline
    \multicolumn{1}{|l|}{\vtop{\hbox{\strut Metric}}} &
    \multicolumn{1}{l|}{\vtop{\hbox{\strut Service}\hbox{\strut type}}} &
    \multicolumn{1}{c|}{MHMP} &
    \multicolumn{1}{c|}{Single-path} &
    \multicolumn{1}{c|}{Path-selection} &
    \multicolumn{1}{c|}{Two-path} \\ \hline
    \multirow{ 2}{*}{Average Latency (ms)} & 1 & 2.95 & 11.11 & 6.01 & 4.11 \\ \cline{2-6}
     & 2 & 1.25 & 3.11  & 1.69 & 1.18 \\ \hline
    \multicolumn{1}{|l|}{Average Power (mW)} & -- & 79.45 & 600.00 & 598.00 & 262.01 \\ \hline
  \end{tabular}}
\end{table}
\endgroup

Table \ref{tab:avgLatPwrS1} summarizes the average latency of the two service types. The multi-path solution achieves the lowest average latency for both service types, while the single-path solutions perform the worst. The two-path splitting and the path-selection baselines lie in between. Note that path-selection can be viewed as a special case of multi-path, where in each slot all power and traffic ratio are assigned to the better path. Therefore, under identical power and channel conditions, the optimal multi-path solution is never worse than path-selection in terms of both average latency.The Table \ref{tab:avgLatPwrS1} also shows that the power of multi-path under fixed latency is lower than which of other paths.Fig. \ref{fig:2} details the performance of the proposed multi-relay, multi-path scheme.Fig. \ref{fig:2}(a) and (b) show Instant-latency CDFs for traffic 1 and 2: the single-path curve shows a long tail (deep fades on a fixed route inflate instantaneous and mean latency); path selection lowers latency by choosing the better one; multi-path exploits both links and yields the leftmost CDF with the shortest tail. Fig. \ref{fig:2}(c) shows the optimized the E2E traffic ratios: with similar average pathloss across the four main routes, per-path trajectories largely overlap (near-balanced splitting); when a sub-path degrades, proportions promptly shift to the others.Fig. \ref{fig:2}(d) shows the optimized per-block transmit power: multi-path~2 consumes slightly more average power than multi-path~1.This indicates better radio quality on Path~2, so the algorithm allocates more power to reduce E2E latency.

We further studied low power, which compares the power of multi-path and baseline with a fixed latency budget of 30ms. As shown in Fig. \ref{fig:3}(a) The low power (LP) scheme’s instant-latency CDF is left-shifted relative to the single-path full-power baseline and shows a much shorter tail, while meeting the latency budget. Fig.\ref{fig:3}(b),(c) and (d) show  the comparison of LP MHMP with LP PS and LP SP respectively. Among these three comparisons, the LP MHMP utilizes routing diversity and block by block power control to meet the delay target, significantly reducing the average transmit power per block and making multi-hop multi-path the main low-power design.

\begin{figure}[h]
\begin{minipage}[t]{0.48\linewidth}
\centering
 \subfigure[Latency savings vs Number of hops.]{\includegraphics[width=1.7in]{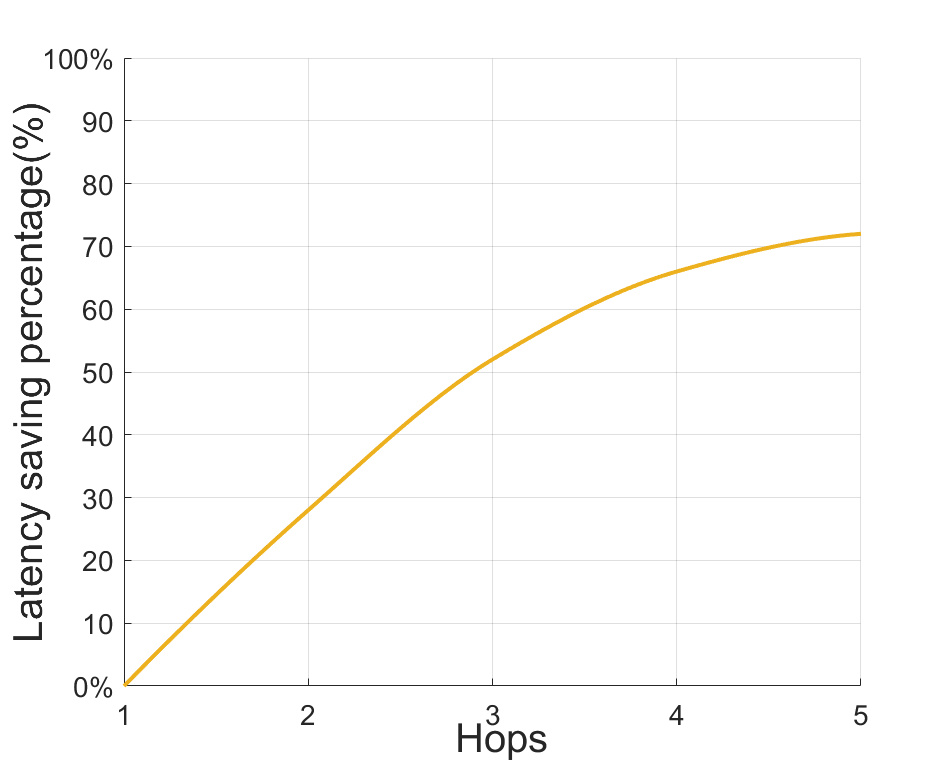}}
\label{fig:4a}
\end{minipage}
\begin{minipage}[t]{0.24\linewidth}
\centering
\subfigure[Power savings vs Number of hops.]{
\includegraphics[width=1.7in]{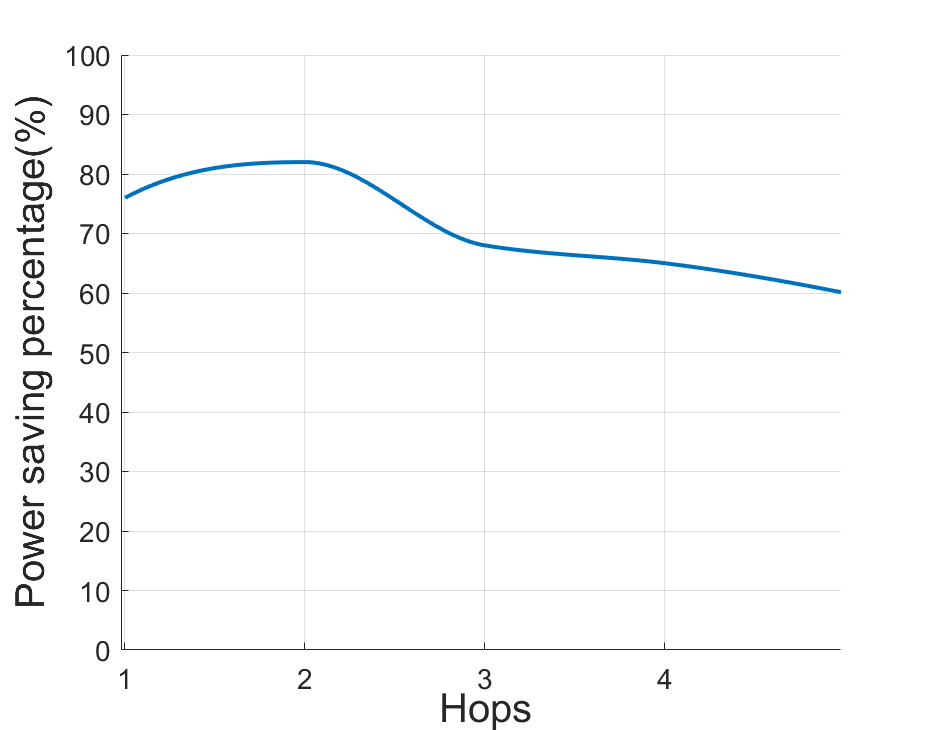}}
\label{fig:4b}
\end{minipage}%
\caption{Impact of Number of Hops on Latency Savings and Power Savings.}
\label{fig:4}
\end{figure}

Fig. \ref{fig:4}(a) shows the variation of delay saving rate of LL MHMP with the number of hops. The results indicate that the saving rate increases monotonically with the number of hops.The reason is that increasing the number of hops expands the available end-to-end paths according to the combination scale. LL MHMP can perform continuous splitting and power joint optimization on multiple paths to avoid deep fading and bottlenecks. Although the increase in resource coupling and path correlation with the deepening of topology leads to a slight slowdown in revenue as the number of hops increases, LL is still able to lower average delay by dynamically rebalancing traffic and power.Afterwards, we investigate  the effect of different layers on power.As shown in Fig. \ref{fig:4}(b), under the same end-to-end latency constraint latency budget, the optimization effect of the proposed LP MHMP exhibits a pattern of first increasing and then decreasing. Increasing the hops from 1 to 2 enlarges the set of available end-to-end routes and thus the degrees of freedom for traffic splitting and power allocation,which reduces the required transmit power and yields a higher saving ratio.More hops imply more hops per end-to-end path.To satisfy the same latency budget, each hop must deliver a higher service rate within a shorter effective time budget, however attaining a higher rate requires substantially larger transmit power, leading to a higher total power and, consequently, a reduced relative saving.

\subsection{Discussion and Advanced Model}
\vspace{-0.15cm}
Algorithm 1 shows the Adaptive Low-Latency/Low-Power Algorithm. The Dynamic power (${\Delta_P}$) and dynamic latency (${\Delta_L}$) depend on QoS performance, user requirements, and vendors solution At the beginning of each block, the scheduler observes channel state information, per-link rates, queue length, and tail-latency, and then switches on demand between two convex subproblems: if latency pressure or backlog is high, it enters Mode LL; otherwise it enters Mode LP. In the selected mode, it jointly optimizes traffic splitting across the available end-to-end paths and the transmit power on each link, subject to capacity, flow-conservation, and power-budget constraints.

\begin{algorithm}[t]
\caption{{ALLP: Adaptive LLP MHMP Scheduler}}
\label{alg:alp-dms-slim}
\SetAlFnt{\footnotesize}
\setlength{\algomargin}{4pt}
\SetInd{0.35em}{0.35em}
\DontPrintSemicolon
\SetKwInOut{Input}{Input}
\Input{$L_{q}^{max},\,P_{tot},\,\Delta_P,\,\Delta_L$}
$P_{i,j,k}(n)\!\leftarrow\!P_{tot}$;\ \textbf{Mode}$\!\leftarrow\!\mathbf{P}$\;
\While{block $k$}{
  measure $\{L_{i,j,k,q}(n)\},D_{i,j,k}$; compute $W_k$\;
  \textbf{Mode}$\!\leftarrow$ \uIf{$W_k>\delta$}{$\mathbf{LL}$}\Else{$\mathbf{LP}$}\;
  \uIf{\textbf{Mode}$=\mathbf{LL}$}{ 
    $P_{i,j,k}(n)\!\leftarrow\!\min\{P_{i,j,k}(n),P_{tot}\}$\;
    solve $L_q(n)$; 
    s.t. capacity/flow/power\;
    \If{infeasible}{$P_{i,j,k}(n)\!\leftarrow\!\min\{P_{tot}{+}\Delta_P,P_{tot}\}$; re-solve}
  }
  \Else{ 
    solve $P(n)$; 
    s.t.  capacity/flow/power\;
    \If{infeasible}{$L_{q}^{max}\!\leftarrow\!\min\{L_{q}^{max}{+}\Delta_L,L_{q}^{max}\}$; re-solve}
  }
  apply $(\{\alpha_{i,j,k,q}(n),P_{i,j,k}(n))$\;
}
\end{algorithm}

\begin{table}[tp]
\caption{Simulation Results}
\label{tb:simu_out}
\centering
\begin{tabular}{|c|c|c|}
\hline
\textbf{Model} & \textbf{Latency(ms)} & \textbf{Power(mW)} \\ \hline
LL & 10.72  &  200 \\ \hline
LP  & 30  & 142.15 \\ \hline
ALLP & 17.56 & 176.82 \\ \hline
\end{tabular}
\end{table}

%% file: conclusion.tex
\section{Conclusion}
\label{sec:con}
In this paper, we proposed a multi-hop multi-path architecture for
multi-type services of traffic towards 6G V2X. We formulated two complementary problems aiming to optimize the E2E latency and the total power of the vehicular network then compared their performance with the proposed baselines in terms of the instant E2E latency per block, and the average E2E latency and the average power over all blocks. We also designed an adaptive LLP scheduler that switches between a fixed-power min-latency mode and a fixed-latency min-power mode to satisfy different QoS requirements. Future work will explore the integration of our proposed framework with the intelligent reflecting surface (IRS) to see the potential benefit of latency and power.

\appendix
\appendices
\subsection{Appendix: Convexity of Problems 1 and 2}

For each hop $(i,j,k,n)$, the transmission rate is
\begin{equation}
D_{i,j,k}(P)=B_{i,j}\log_{2}\!\left(1+\frac{P-P^{\mathrm{Loss}}_{i,j,k}}{B_{i,j}N_{0}}\right).
\end{equation}
Since $\log(1+x)$ is concave for $x>-1$, it follows that
\begin{equation}
D_{i,j,k}(P) \text{ is concave and } D_{i,j,k}(P)>0.
\label{eq:concave_rate}
\end{equation}

Let
\begin{equation}
A_{i,j,k,q}(\alpha)=\big(\alpha_{i,j,k,q}\lambda_{i,k,q}+n_{i,k,q}\big)M_{i,k,q},
\end{equation}
which is affine in $\alpha_{i,j,k,q}$. The hop–latency term is
\begin{equation}
L_{i,j,k,q}(\alpha,P)=\frac{A_{i,j,k,q}(\alpha)}{D_{i,j,k}(P)}.
\end{equation}

Consider its epigraph:
\begin{equation}
\operatorname{epi}(L)
=\left\{(\alpha,P,t): A_{i,j,k,q}(\alpha)\le t\,D_{i,j,k}(P),\ t\ge 0\right\}.
\label{eq:epi}
\end{equation}
Since $D_{i,j,k}(P)$ is concave and positive, the hypograph $\{(P,r): r\le D_{i,j,k}(P)\}$ is convex, and the inequality $A(\alpha)\le t r$ is linear in $(\alpha,t,r)$. Hence, \eqref{eq:epi} is a convex set, which implies
\begin{equation}
L_{i,j,k,q}(\alpha,P) \text{ is jointly convex in } (\alpha,P).
\end{equation}

The E2E latency satisfies
\begin{equation}
u_{b,q}(\alpha,P)=\sum_{(i,j,k)\in \mathcal{P}_b} L_{i,j,k,q}(\alpha,P),
\end{equation}
and therefore
\begin{equation}
u_{b,q}(\alpha,P) \text{ is convex.}
\end{equation}

The objective of Problem 1 is
\begin{equation}
\min_{\alpha,P} \ \max_{q} u_{b,q}(\alpha,P),
\end{equation}
and $\max_{q}(\cdot)$ preserves convexity. Constraints C1–C5 are affine (or convex upper bounds $u_{b,q}\le L_{q}^{\max}$). Therefore, Problem 1 is convex.

For Problem 2, the objective is
\begin{equation}
\min_{\alpha,P} \ \sum_{n,i,j,k} P_{i,j,k}(n),
\end{equation}
which is linear in $P$, and it is subject to the same convex latency expression and affine constraints. Hence, Problem 2 is also convex.